\documentclass[%
 reprint,
showpacs,preprintnumbers,
 amsmath,amssymb,
 aps,
]{revtex4-1}

\usepackage{graphicx}
\usepackage{dcolumn}
\usepackage{bm}

\begin{document}

\title{The effect of $SU(4)$ group factor in the thick center vortex potentials}

\author{Shahnoosh Rafibakhsh}

\email{rafibakhsh@ut.ac.ir}
\affiliation{
 Plasma Physics Research Center, Science and Research Branch, Islamic Azad University, Tehran, Iran
}

\date{\today}

\begin{abstract}
The group factor for the higher representations of the $SU(4)$ gauge group is investigated. For the zero N-ality representations, the minimum values of the real part of the group factor which appear at large distances, are located at the points where $50\%$ of the maximum vortex flux is in the Wilson loop. This amount is equal to $60\%$ for the representations which belong to N-ality=$1$ class except the fundamental representation. The value of the real part of the group factor at the minimum points is equal to the value of center of the group factor at intermediate distances. The appearance of these points is due to the difference between the asymptotic and intermediate string tensions. The group factor passes the phase factor to create the larger intermediate string tension than the asymptotic one.
\end{abstract}



\maketitle


\section{\label{sec:1}Introduction}	

Today everyone knows that Quantum Chromodynamics (QCD) is a promising theory of strong interactions. In the low energy regime of this theory, there are still open unsolved questions like quark confinement, which has been a challenging problem over thirty years. A lot of phenomenological models have been introduced to determine the behavior of QCD in this regime. In these models, it is believed that the QCD vacuum is responsible for confinement. Among wide variety of phenomenological models, center vortex theory based on magnetic degrees of freedom gives an acceptable explanation of quark confinement \cite{thooft, *vin, *cornwall, *feynman, *nielsen, *ambjorn1, *ambjorn2, *ambjorn3, *olesen}. Numerical lattice calculations show the evidence for the existence of the topological objects called center vortices in the vacuum \cite{deldebbio,*deldebbio2,*tomboulis,*tomboulis2,*reinhardt}. Center vortices are quantized magnetic flux tubes (or surfaces) which are closed according to the Bianchi identity \cite{dirac}. They are created by a singular gauge transformation and have finite energy per unit length (or surface). 

Based on the center vortex theory, the area law fall-off for the Wilson loop which leads to quark confinement is due to the quantum fluctuations of the vortices interacting with the Wilson loop. This interaction affects the Wilson loop in the representation $r$ by an element of the $SU(N)$ gauge group center:
\begin{equation}
W_r(C)\longrightarrow z^{k_r} W_r(C),
\label{W}
\end{equation}
where $k_r$ is the N-ality of the representation $r$. On the other hand, one might say that when the vortex does not link with the Wilson loop, the loop remains unaffected. 

Although the confinement at large distances is produced by these types of vortices, they cannot give the linear potential at intermediate distance. To get the intermediate potentials, Faber {\it et al.} generalized this model to thick center vortex model to eliminate this defect \cite{faber}. One might claim that the thick center vortex model explains the following features of confinement very well:\\
\begin{enumerate}
\item The coulombic potential at short distances that has been proved in two different methods: Deldar {\it et al.} showed by increasing the role of vortex fluxes piercing Wilson loops with contributions close to the trivial center element and by fluctuating the vortex core size, the coulombic potential is produced \cite{deldar2009}. In a different work, the fluctuation of non-quantized, closed magnetic flux lines has been considered by Faber {\it et al.} to calculate the short range potential \cite{denis}.
\item Casimir Scaling of string tensions extended from the onset of confinement to the onset of screening which has been proved precisely by lattice calculations \cite{deldarlat,bali}. This theory claims that string tensions are roughly proportional to the eigenvalues of the quadratic Casimir operators. The results obtained from the thick center vortex model show that the string tension of different representations is qualitatively in agreement with Casimir scaling \cite{faber,greensite, deldar-jhep,deldar2005,deldar2007,deldar2009,deldar2010}, even for exceptional G(2) gauge group \cite{deldar-g2}.
\item N-ality dependence of asymptotic string tensions at large distances which frames that the asymptotic string tension of each representation is the same as that of the lowest dimensional representation with the same N-ality. Calculations in thick center vortex model confirm this feature of confinement \cite{faber,greensite, deldar-jhep,deldar2005,deldar2007,deldar2009,deldar2010}.
\end{enumerate}

When the vortex is thickened, the vortex core may overlap the perimeter of the Wilson loop and a part of the vortex flux enters the loop. Therefore, the role of the center element is replaced by the group factor which interpolates between a center element -when the vortex core is completely inside the Wilson loop- and $1$ -when the  Wilson loop and the vortex do not interact. Therefore, studying the behavior of the group factor at different quark separation distances will give useful information about the potential between quark and anti-quark.

In the next section, the thick center vortex model is studied and it is explained how the group factor appears in the model. Moreover, the maximum flux of the $SU(4)$ vortices is calculated and it is discussed which of the Cartan generators should be used in the calculation. The behavior of the group factor at different quark separation is discussed in section three and the results are given in section four.

\section{\label{sec:2}The group factor in the thick center vortex model}

In the original vortex model, center elements of the gauge group are responsible for confinement. For thin vortices two scenarios may occur:\\ 
1) Vortices and the Wilson loop do not link and therefore the Wilson loop is unaffected:
\begin{equation}
W(C)=Tr \big[U...U\big]\longrightarrow Tr \big[U...I...U\big].
\label{W1}
\end{equation}
2) Vortices may pierce the minimal area of the Wilson loop. The effect of this interaction is inserting a center element $z$ between link operators:
\begin{equation}
W(C)=Tr \big[U...U\big]\longrightarrow Tr \big[U...z...U\big].
\label{W2}
\end{equation}
Therefore, if {\it $f$} is the probability of piercing a plaquete by the $z$ vortex of the $SU(2)$ gauge group, the Wilson loop might be written as:
\begin{eqnarray}
<W(C)>=\prod_{x\in A}{[(1-f)+f(-I)]}<W_0(C)> \nonumber\\
=\exp \big[-\sigma(C)A\big]<W_0(C)>,
\label{<W>}
\end{eqnarray}
where $<W_0(C)>$ is the Wilson loop expectation value when no vortices pierce the minimal area of the loop and the string tension is
\begin{equation}
\sigma=-\ln(1-2f).
\label{sigma}
\end{equation} 
Then the thin vortices give the correct N-ality dependence of the potentials at large distances while the intermediate linear potential is lost. Thus, to get the intermediate potentials, the vortices are thickened. Thickening the vortices leads to a new scenario. The vortex may overlap the perimeter of the Wilson loop and one has to consider a distribution for the flux carried by the vortex. In this case, a part of the vortex flux might enter the Wilson loop and the center element is replaced by a group element $G$ which is a unitary matrix called the $SU(N)$ group factor. This factor parametrizes the influence of the vortex on the Wilson loop:
\begin{equation}
W(C)=Tr \big[UU...U\big]\longrightarrow Tr\big[UU...G...U\big],
\label{W3}
\end{equation}
where
\begin{equation}
G(x,S)=S \exp \big[i \vec{\alpha}^n_c.\vec{H}\big]S^\dagger,
\label{G}
\end{equation}
where the $H_i$'s, $\big\lbrace i=1,...,N-1 \big \rbrace $ are the generators spanning the Cartan sub-algebra and $S$ is an $SU(N)$ group element in representation $r$. $\alpha^{(n)}_c$ represents the flux distribution of the $z_n$ vortex. In general, in an $SU(N)$ gauge group, there are $N-1$ types of center vortices corresponding to the number of the center elements of the gauge group.

Based on the assumptions of the model, the random group orientations associated with $S_i$ are considered uncorrelated. Therefore, one has to average $G$ over all orientations in the group manifold: 
\begin{eqnarray}
\bar{G}(\vec{\alpha})\equiv {\cal G}_r \big[\vec{\alpha}^{(n)} \big]I=\int dS S \exp\big[i\vec{\alpha}.\vec{H}\big]S^\dagger \nonumber\\
=\frac{1}{d_r}Tr \exp\big[i\vec{\alpha}^{(n)}.\vec{H}\big],
\label{gr}
\end{eqnarray}
where $d_r$ is the dimension of representation $r$. Therefore, the potential energy between static sources induced by the vortices is 
\begin{equation}
V(R) = -\sum_{x \in A}\ln\left\{ 1 - \sum^{N-1}_{n=1} f_{n}
(1 - {\mathrm {Re}} {\cal G}_{r} [\vec{\alpha}^n_{C}(x)])\right\}.
\label{V}
\end{equation}
In Eq.~(\ref{V}), it is observed that the factor ${\cal G}_r \big[\vec{\alpha} \big]$ owns an important role in producing the potentials. According to Eq.~(\ref{gr}), this factor depends on the flux $\alpha_c(x)$. Furthermore, the vortex profile $\alpha_c(x)$ depends on what fraction of the vortex flux enters the loop $C$. Therefore, it depends on both the shape of the loop $C$, and the position $x$ of the center of the vortex core, relative to the perimeter of the loop, as the following:\\
\begin{enumerate}
\item When the vortex core is entirely outside the planar area enclosed by the Wilson loop, it cannot affect the loop: 
\begin{equation}
\exp \big[i \vec{\alpha}^{(n)}.\vec{H}\big]=I\Longrightarrow {\vec{\alpha}}^{(n)}=0.
\label{alpha_min}
\end{equation}
In this case, the lower limit for $\alpha$ is achieved.\\
\item When the vortex core is completely inside the Wilson loop area, then the influence of the vortex on the Wilson loop is given by a center element:
\begin{equation}
\exp\big[i \vec{\alpha}^{(n)}.\vec{H}\big]=z_n I \quad \Longrightarrow {\vec{\alpha}}^{(n)}={\vec{\alpha}}^{(n)}_{\mathrm max},
\label{alpha-max}
\end{equation}
where $I$ is the unit matrix. ${\alpha}^{(n)}_{max}$ is the upper limit of the vortex profile and should be calculated for every vortex type in the $SU(N)$ gauge group. \\
\item As $R\rightarrow 0$ then $\alpha \rightarrow 0$.
\end{enumerate}

It should be noticed that all choices for $\alpha (x)$ must lead to a well-defined potential which means both Casimir proportionality and N-ality dependence must be preserved. The vortex profiles checked in this paper are listed below:
\begin{enumerate}
\item The flux introduced by Faber {\it et al.} with two free parameters $a$ and $b$ is \cite{faber}:
\begin{equation}
\alpha(x)=\frac{\alpha_{\mathrm {max}}}{2}(1-\tanh(ay(x)+\frac{b}{R})),
\label{tanh}
\end{equation}
where $R$ is the space-extent of the Wilson loop with finite time-extent and $x$ represents the position of the vortex core. $a$ is proportional to the inverse of the vortex thickness and the parameter $b$ introduces a dependency on the space extent $R$ of the Wilson loop into the vortex profile. $y(x)$ is
\begin{equation}
y(x)=\begin{cases}
x-R \quad\mathrm{for}\quad|R-x|\le|x|\\
-x ~~~\quad\mathrm{for}\quad|R-x|>  |x|.
\end{cases}
\end{equation}
\item A similar profile with only one parameter $\acute{a}$ which corresponds to the inverse of the vortex thickness, was introduced by Faber {\it et al.} \cite{denis}:
\begin{equation}
\alpha(x)=\frac{\alpha_{\mathrm {max}}}{2}(\tanh({\acute{a}}(x+\frac{R}{2}))-\tanh({\acute{a}}(x-\frac{R}{2})).
\label{newtanh}
\end{equation}
The parameter $b$ has been removed because lattice calculations do not show the influence of the Wilson loop on the vortices. It should be noticed that $\acute{a}$ is not the same as $a$ in Eq.~(\ref{tanh}).
\item The following profile was introduced by Deldar in Ref.~\cite{deldar-jhep}:
\begin{equation}
\alpha (x)=\beta (x)-\beta (x-R),
\label{newflux}
\end{equation}
where $\beta (x)$ -the amount of the vortex flux contained in different regions- is given by:
\begin{equation}
\beta(x)=
\begin{cases}
 \frac{\alpha_{\mathrm {max}}}{2} & \quad x\geq a\\
 \frac{\alpha_{\mathrm {max}}}{2}\big(1-\exp[b(1-\frac{1}{(\frac{x}{a}-1)^2})]\big) & \quad 0\le x \le a\\
-\frac{\alpha_{\mathrm {max}}}{2}\big(1-\exp[b(1-\frac{1}{(\frac{x}{a}+1)^2})]\big) & \quad -a\le x \le 0\\
-\frac{\alpha_{\mathrm {max}}}{2} & \quad x\leq -a.
\end{cases}
\label{beta}
\end{equation}
By varying the free parameters $a$ and $b$, the shape of the profile changes. For instance, $a=150$ and $b=10$ lead to a profile similar to Eq.~(\ref{tanh}) while changing the parameter $b$ to $0.01$ turns the profile to a delta function which violates Casimir scaling.
\end{enumerate}

In Sec.~\ref{sub:1}, $\alpha_{\mathrm{max}}$ for the two vortices of the $SU(4)$ gauge group -using the upper limit criterion- is calculated as shown in Eq.~(\ref{alpha-max}).

\subsection{\label{sub:1}Calculating the maximum flux}

\begin{figure}
\includegraphics[scale=0.7]{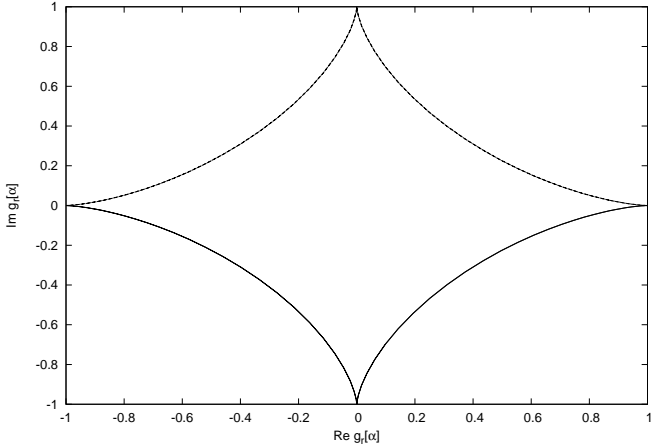}
\caption{The imaginary part of ${\cal G}_r[\alpha]$ versus the real part in the $SU(4)$ gauge group for $z_1=\exp(\frac{\pi i}{2})$, $z_2=\exp(\pi i)$, $z_3=\exp(\frac{3\pi i}{2})$ vortices and the vacuum trivial structure. In the calculation of $\mathrm{Re}{\cal G}_r[\alpha]$ and $\mathrm{Im}{\cal G}_r[\alpha]$, the vortex flux given in Eq.~(\ref{tanh}) with $a=0.05$ and $b=4$ and the Cartan generator $H_3$ have been used. It is observed that in case of using only $H_3$, the group factor can produce the four center elements of this gauge group- including the trivial element.}
\label{fig:trace}
\end{figure}

As mentioned in Sec.~\ref{sec:2}, there are $N-1$ types of vortices in an $SU(N)$ gauge group, but not all of them are independent. In fact, the vortices of types $n$ and $N-n$ are conjugate and the fluxes carried by them are in the opposite directions. So, in the $SU(4)$ gauge group, two types of center vortices fill the QCD vacuum. Now from Eq.~(\ref{alpha-max}), the maximum flux of the two vortices of this gauge group can be calculated. For the $z_{1}=\exp(\frac{\pi i}{2})$ vortex of the $SU(4)$, one may write:
\begin{equation}
\exp\big[i \vec{\alpha}^{(1)}.\vec{H}\big] =\exp (\frac{\pi i}{2})I.
\label{grmax-su4}
\end{equation}
The cartan generators of the $SU(4)$ are as the following:
\begin{eqnarray}
H_1&=&\frac{1}{2} \mathrm{diag}(1,-1,0,0), \\
H_2&=&\frac{1}{2\sqrt{3}}\mathrm{diag}(1,1,-2,0),\\
H_3&=&\frac{1}{2\sqrt{6}}\mathrm{diag}(1,1,1,-3).
\label{H-su4}
\end{eqnarray}
If the three of the above matrices are used, the calculations go as follows:
\begin{equation}
\exp \big[i{\alpha}_1^{(1)} H_1 + i{\alpha}^{(1)}_2 H_2 + i{\alpha}^{(1)}_3 H_3\big]=\exp(\frac{\pi i}{2})I,
\label{alpha-su4}
\end{equation}
where upper index $(1)$ represents the vortex type and lower indices imply the projection of $\vec{\alpha}_{\mathrm{max}}$ on the directions of the cartan generators. It should be noticed that the $\mathrm{max}$ indices for the projection parts have been omitted. Therefore,
\begin{equation}
{\alpha}_1^{(1)}=-2\pi \quad , \quad {\alpha}_2^{(1)}=-\frac{2\pi}{\sqrt{3}} \quad , \quad {\alpha}_3^{(1)}=-\frac{2\pi}{\sqrt{6}}.
\label{all-su4-max}
\end{equation}
Another possible way is to use only $H_3$ matrix:
\begin{equation}
\exp \big[i{\alpha}^{(1)}_{\mathrm{max}} H_3 \big]=\exp(\frac{\pi i}{2})I\Longrightarrow {\alpha}^{(1)}_{\mathrm{max}}=\pi\sqrt{6}.
\label{z1-max}
\end{equation}
Both of the above normalization methods lead to the same result as the factor ${\cal G}_r[\vec{\alpha}]$ includes all the center elements of the gauge group in both cases (see Fig.~\ref{fig:trace}). However, if one uses only $H_1$ or $H_2$ in the calculations, it is impossible to get all the center elements.
The same method could be used to calculate the maximum flux of the $z_2$ vortex. I recall that using only $H_3$ in the calculation gives the same result as using the three diagonal generators of the $SU(4)$. Thus, the second normalization method given in Eq.~(\ref{z1-max}) is used:
\begin{equation}
\exp \big[i{\alpha}^{(2)}_{\mathrm{max}} H_3 \big]=\exp(\pi i)I\Longrightarrow {\alpha}^{(2)}_{\mathrm{max}}=2 \pi \sqrt{6}.
\label{z2-max}
\end{equation}
 
Now Cartan generators should be calculated for the higher representations which is done in the next section.

\subsection{\label{sub:2}Higher representations of the $SU(4)$ gauge group}

To calculate ${\cal G}_r[\vec \alpha]$ for higher representations, the corresponding $H_3$ matrix should be obtained from \cite{georgi}
\begin{equation}
{\big(H_3^{D_1\otimes D_2}\big)}_{ix,iy}=\big (H_3^{D_1}\big) \delta_{xy}+\delta_{ij} \big (H_3^{D_2}\big),
\label{higher}
\end{equation}
where $H_3^a$s are the $SU(4)$ generators for $D_1\otimes D_2$, $D_1$ and $D_2$ representations, respectively. If ${X_r^i;i=1,2,...,d_r}$ is considered as the vector basis for the representation $d_r$, the elements of $H_3^a$ could be found by:
\begin{equation}
H_3^rX_r^i=\sum_{j=1}^{d_r} C_{ij} X_r^j.
\label{H-high}
\end{equation}
The aim of this paper is to investigate the behavior of the group factor for higher representations. Now we find a few representations which were discussed in the previous works \cite{deldar2005, deldar2007, deldar2009}. Then by finding the basis vectors of them, the $H_3$ generator of the higher representations is obtained. 
\begin{equation}
4 \otimes \bar4=\bm {15} \oplus 1,
\label{adjoint}
\end{equation}
\begin{equation}
4 \otimes 4=\bm {10} \oplus \bm {6},
\label{diquark}
\end{equation}
\begin{equation}
4 \otimes 4 \otimes 4=\bm {20_s} \oplus 20_m \oplus 20_m \oplus \bar4,
\label{20}
\end{equation}
\begin{equation}
4 \otimes 4 \otimes 4 \otimes 4=3 \times {45} \oplus \bm {35_s} \oplus 2 \times 20_{\Box} \oplus 3 \times 15 \oplus 1.
\label{35}
\end{equation}
It should be mentioned that the behavior of the representations written in bold is investigated in this paper. If $v^i$ and $v_j$ are considered as the basis vectors for representations $4$ and $\bar 4$, respectively, the basis vectors for the higher representations -mentioned above- could be calculated from tensor calculus. More details for the basis vectors of each representation are given in the appendix. From Eq.~(\ref{H-high}) and the basis vectors, the $H_3$ generator for the above representations is calculated.\\
Representation $6$:
\begin{equation}
H_3^6=\frac{1}{\sqrt6}diag\big(1,1,-1,1,-1,-1\big),
\label{H6}
\end{equation}
Representation $10$:
\begin{equation}
H_3^{10}=\frac{1}{\sqrt6}diag\big(1,1,1,-1,1,1,-1,1,-1,-3\big),
\label{H10}
\end{equation}
Representation $15$:
\begin{equation}
H_3^{15}=\frac{2}{\sqrt6}diag\big(0,0,0,1,0,0,0,1,0,0,0,1,-1,-1,-1\big),
\label{H15}
\end{equation}
Representation $20_s$:
\begin{widetext}
\begin{equation}
H_3^{20_s}=\frac{1}{2\sqrt6}diag\big(3,3,3,-1,3,3,-1,3,-1,3,3,-1,3,-1,-5,3,-1,-5,-5,-9\big),
\label{H20}
\end{equation}
\end{widetext}
Representation $35_s$:
\begin{widetext}
\begin{equation}
H_3^{35_s}=\frac{2}{\sqrt6}diag\big(1,1,1,0,1,1,0,1,0,-1,1,1,0,1,0,-1,1,0,-1,-2,1,1,0,1,0,-1,1,0,-1,-2,1,0,-1,-2,-3\big).
\label{H35}
\end{equation}
\end{widetext}
In Sec.~\ref{results}, the plots of $Re{\cal G}_r(\vec{\alpha})$ are investigated for different quark separations and the place of the minimum points observed in the plots is discussed.

\section{\label{results}Results and discussions}
As it was mentioned previously, ${\cal G}_r(\vec{\alpha})$ plays an important role in the potential between quarks. This factor changes between $1$ and $\exp (\frac{2 \pi i n k}{N})$ corresponding to the type of vortex $n$ and N-ality=$k$ of the representation. Depending on the representations, the way ${\cal G}_r(\vec{\alpha})$ changes between these two limits differs which means in some representations such as $15$, $20_s$ and $35_s$, the group factor passes the phase factor. 

\begin{figure}
\includegraphics[scale=0.7]{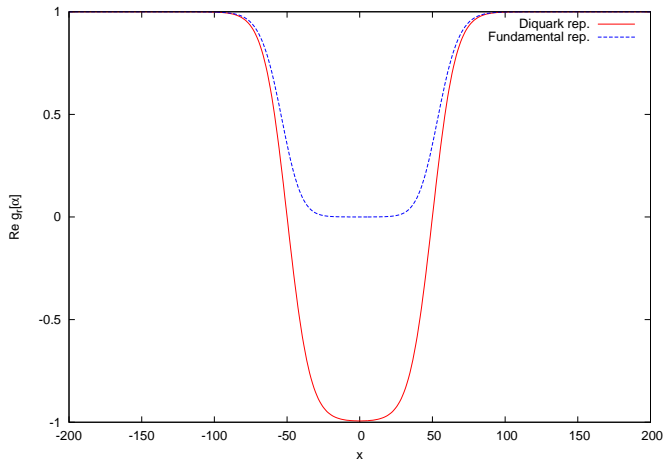}
\caption{\label{fig:both} $Re{\cal G}_r({\alpha})$ versus the position of the vortex core $x$ for the $z_1$ vortex of the $SU(4)$ and for diquark and fundamental representations of this gauge group using the flux given in Eq.~(\ref{newtanh}) with $\acute {a}=0.04$ and $R=100$. The vortex thickness is equal to $25$ so when $x=0$, the vortex is completely inside the Wilson loop. As it is expected, the group factor changes between $1$ and $0$ for the fundamental representation and between $1$ and $-1$ for the diquark one. Therefore, for these representations $Re{\cal G}_r({\alpha})$ behaves normally.}
\end{figure}

Fig.~\ref{fig:both} shows the real part of the $SU(4)$ group factor versus the position of the $z_1$ vortex core for the fundamental and diquark representations with $4$-ality equal to $1$ and $2$, respectively. The Wilson loop legs are located at $x=-50$ and $x=50$ so $R=100$. The flux used to produce this figure is the same as in Eq.~(\ref{newtanh}) and the parameter $\acute{a}$ is chosen equal to $0.04$, hence the vortex thickness -which is proportional to $\frac{1}{a}$- is equal to 25. As we expect, $Re{\cal G}_r({\alpha})$ varies between 1 -when there is no overlap between the vortex and the Wilson loop- and the real part of the phase factor $\exp (\frac{ \pi i k}{2})$ -when the vortex is entirely enclosed by the Wilson loop. Therefore, $Re{\cal G}_r({\alpha})$ behaves as expected for these two representations. But the results are different for higher representations.

\begin{figure}
\includegraphics[scale=0.7]{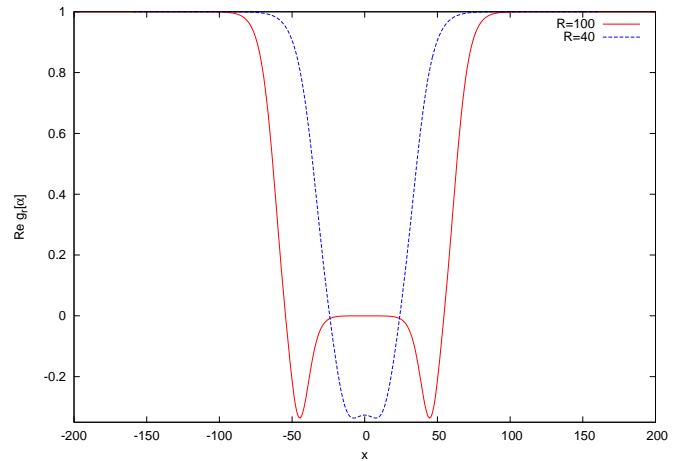}
\caption{\label{fig:gr20} $Re{\cal G}_r({\alpha})$ versus the position of the $z_1$ vortex core $x$ for representation $20_s$ and for $R=40$ and $R=100$. The flux used to produce this figure is given in Eq.~(\ref{newtanh}) with $\acute {a}=0.04$. When $R=100$, two minimum values are observed at $x=-45$ and $x=45$ where about $60\%$ of the maximum vortex flux enters the Wilson loop. The value of $Re{\cal G}_r({\alpha})$ at the minimum points is equal to $-0.33$. The vortex thickness is equal to $25$ so when the position of the vortex core is located at $x=0$, the vortex is completely inside the Wilson loop. Thus, $Re{\cal G}_r({\alpha})=1$ and one may conclude that the asymptotic string tension is produced. When the Wilson loop spatial extent is decreased to $40$ and the location of the vortex core is at the middle of the Wilson loop, $Re{\cal G}_r(\vec{\alpha})$ is near $-0.33$. So $(Re{\cal G}_r(\alpha))_{\mathrm {min}}$ at large distances, is equal to the center of the group factor at intermediate distances.}
\end{figure}

Fig.~\ref{fig:gr20} is the same as Fig.~\ref{fig:both} but for representation $20_s$ with $4$-ality=$1$ and for two values of the space extent $R$. The group factor is expected to alter between $1$ and $0$ for $R=100$. But as it is seen in this figure, the two minimum values are observed at $x=-45$ and $x=45$ which have a value equal to $-0.33$. So $Re{\cal G}_r({\alpha})$ has passed $0$ which is the real part of the phase factor for representation $20_s$. Now the influence of the vortex on the Wilson loop is investigated at different positions to find out how the group factor changes. The center vortex theory states that the Wilson loop is unaffected when there is no interaction between the vortex and the Wilson loop hence $Re{\cal G}_r({\alpha})=1$. The effect of the vortex on the Wilson loop starts when it overlaps the perimeter of the loop. If the position of the center of the vortex core is placed at $x=-45$, about $60\%$ of the maximum flux enters the Wilson loop. In this case:
\begin{equation}
Re{\cal G}_r(\vec\alpha)=Re\frac{1}{d_r} Tr \exp\big[\frac{3i}{5}\alpha_{\mathrm max}\times H_3^{(20)} \big]=-0.33,
\label{G-half}
\end{equation}
and the minimum point of the group factor is created. The vortex is completely inside the Wilson loop when the position of the center of the vortex is placed at $x=0$. So according to Eq.~(\ref{W}), the Wilson loop of representation $20_s$ is multiplied by a phase factor $\exp (\frac{\pi i k}{2})$ where $k=1$. Therefore, we expect at this point that $Re{\cal G}_r(\alpha)$ is equal to $0$. It means that the Wilson loop and the vortex are totally linked to each other and the asymptotic string tension is achieved. As the vortex starts leaving the Wilson loop, a part of the vortex flux quits the loop; So that when the position of the vortex core is placed at $x=45$, $60\%$ of the maximum flux carried by the vortex still remains in the Wilson loop. So again the value of $Re{\cal G}_r(\alpha)$ becomes equal to $-0.33$. Now the distance between quark and anti-quark is decreased to produce the intermediate potential. In Fig.~\ref{fig:gr20}, it is seen when $R=40$ and the vortex core is located at $x=0$, which is the middle point of the Wilson loop, the value of the group factor is equal to $-0.33$. At this stage, the intermediate string tension is produced. Strictly speaking, the value of the group factor when the position of the vortex core is located at the middle of the Wilson loop determines the potential behavior and it is called the "center of the group factor" in this paper. Therefore, one may conclude that the minimum values seen for $R=100$ in Fig.~\ref{fig:gr20} are the points which the center of the group factor reaches at intermediate distances.

\begin{figure}
\includegraphics[scale=0.7]{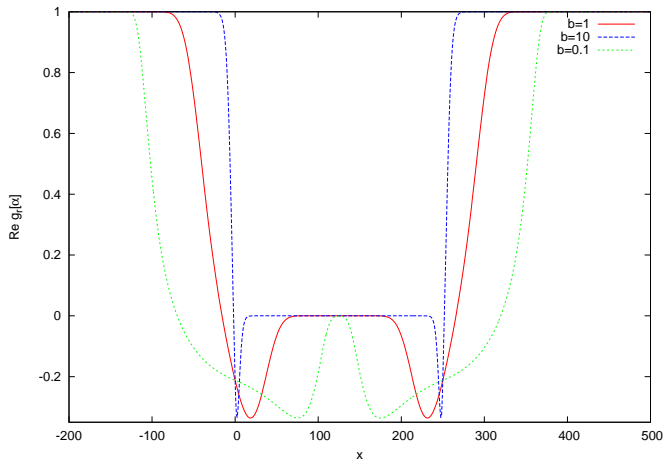}
\caption{\label{fig:new20}The same as Fig.~\ref{fig:gr20} but for the flux given in Eq.~(\ref{newflux}) with $a=150$, $b=10$, $b=1$, $b=0.1$ and $R=250$. The two minimum points are observed in all three curves and $(Re{\cal G}_r({\alpha}))_{\mathrm{min}}=-0.33$. The minimum points are located at the positions where $60\%$ of the maximum flux of the $z_1$ vortex enters the Wilson loop.}
\end{figure}

The most interesting aspect about the group factor is that the value of $Re{\cal G}_r(\alpha)$ at the minimum points is independent of the flux used in the calculation. This issue is understood by a comparison between Fig.~\ref{fig:gr20} and the three curves in Fig.~\ref{fig:new20} where the flux in Eq.~(\ref{newflux}) has been used with $a=150$ and three values for the parameter $b$. The space extent $R$ of the Wilson loop is equal to $250$ so that the center of the group factor reaches $0$ for the three curves. It is observed that value of $(Re{\cal G}_r(\alpha))_{\mathrm {min}}$ remains the same even when the shape and the type of the vortex flux change. The similar result is produced if the flux in Eq.~(\ref{tanh}) is investigated.

\begin{figure}
\includegraphics[scale=0.7]{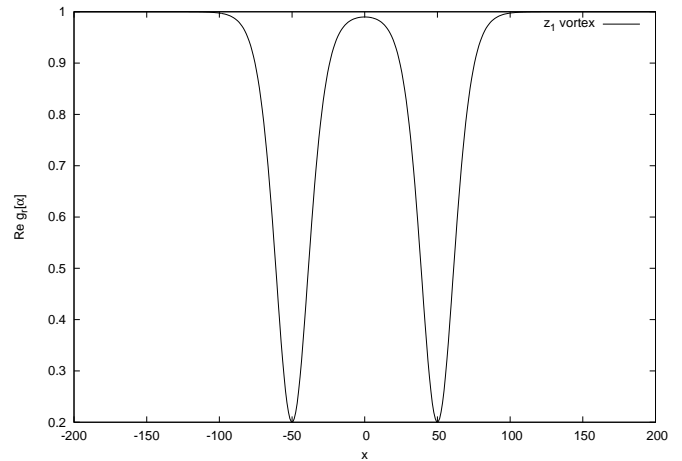}
\caption{\label{fig:gr15} The same as Fig.~\ref{fig:gr20} but for the adjoint representation. Two minimum values are observed at $x=-50$ and $x=50$. The value of $Re{\cal G}_r(\vec{\alpha})$ at the minimum points is equal to $0.2$. These points are the positions where half of the vortex maximum flux enters the Wilson loop. When the position of the vortex core is located at $x=50$, $Re{\cal G}_r({\alpha})=1$ and one may conclude that the asymptotic string tension is produced.}
\end{figure}

\begin{figure}
\includegraphics[scale=0.7]{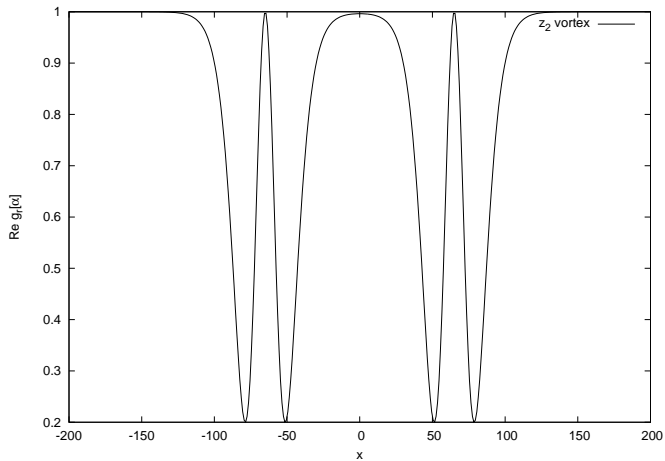}
\caption{The same as Fig.~\ref{fig:gr15} but for the $z_2$ vortex. For $x=-79$ and $x=78$, one fourth of the maximum flux of the $z_2$ vortex is in the Wilson loop which is equal to half of the maximum flux of the $z_1$ vortex. So, as in Fig.~\ref{fig:gr15}, $Re{\cal G}_r(\vec{\alpha})=0.2$ and analogue for $x=-52$ and $x=51$ where the three fourth of the maximum vortex flux are in the loop. The Wilson loop and the vortex are linked completely at $x=0$ and the value of $Re{\cal G}_r(\vec{\alpha})$ is equal to $1$ at this point.}
\label{fig:gr152}
\end{figure}

Another fact about the group factor is that the portion of the vortex flux which enters the Wilson loop at the minimum points, is different for representations with different N-ality. This matter makes us investigate the behavior of $Re{\cal G}_r(\alpha)$ for the adjoint representation with $4$-ality=$0$. The real part of the group factor of the adjoint representation versus $x$ for $R=100$ is observed in Fig.~\ref{fig:gr15} using the same flux as of Fig.~\ref{fig:gr20}. When the $z_1$ vortex core is located at $x=-50$ and $x=50$, half of the maximum flux is in the Wilson loop and at these points $Re{\cal G}_r({\alpha})=0.2$. 
The behavior of $Re{\cal G}_r(\alpha)$ for the $z_2$ vortex can be analyzed in the similar way. Fig.~\ref{fig:gr152} is the same as Fig.~\ref{fig:gr15} but for the $z_2$ vortex. In this figure, $R$ has changed to $130$ so that the center of the group factor reaches $1$ for the $z_2$ vortex which contains two $z_1$ vortices. When the position of the $z_2$ vortex core is placed at $x=-79$ and $x=78$, one fourth of the $z_2$ vortex flux is in the Wilson loop. This amount of flux is equivalent to half of the maximum flux carried by one $z_1$ vortex. Thus one can expect -from the above discussion for the $z_1$ vortex- that the amount of $Re{\cal G}_r(\alpha)$ is equal to $0.2$. At $x=-65$ and $x=64$, half of the $z_2$ vortex flux or equivalently the total flux of one $z_1$ vortex is in the Wilson loop. Therefore, $Re{\cal G}_r(\alpha)$ reaches $1$. When $x=-52$ and $x=51$, three fourth of the $z_2$ vortex enter the Wilson loop which is equal to half of the flux carried by one $z_1$ vortex and $Re{\cal G}_r(\alpha)$ equals to $0.2$. The $z_2$ vortex is completely inside the loop when $x=0$ and the Wilson loop changes by a phase factor $(\exp \pi i k)$ with $k=0$ hence $Re{\cal G}_r(\alpha)$ becomes equal to $1$. The behavior of the group factor of representation $35_s$ is like as of the adjoint one. For this representation, ${(Re{\cal G}_r(\alpha))}_{\mathrm min}=-0.25$. Thus, one might claim that the number of the quarks and anti-quarks which make the representation changes the portion of the remaining flux inside the Wilson loop at the minimum points. For more investigation, we look at Eqs.~(\ref{adjoint}), (\ref{20}) and (\ref{35}). It is observed that representation $20_s$ is built from three quarks while four quarks form representations $35_s$. Also, one quark and one anti-quark make the adjoint representation. On the other hand, for zero $4$-ality representations built from even numbers of quarks and anti-quarks, the minimum points in the plots are the places where half of the maximum vortex flux is in the Wilson loop while for higher representations which are produced from an odd number of quarks, the minimum points are located at the positions where $60\%$ of the maximum flux is in the loop. One may find the reason of this behavior in the difference between the slope of the potentials at intermediate and large distances.

In our previous works \cite{deldar2005, deldar2007}, it was shown  that at intermediate distances, the ratio of the string tension of each representation to that of the fundamental one is proportional to Casimir scaling. At large distances, the representations with the same $4$-ality get the same slope. The first row of Tab.~\ref{tab} shows the representations along with the corresponding Casimir scaling in the second row. The third row represents the $4$-ality of each representation. Thus, for representations $15$ and $35_s$ screening is observed. The potential of representation $20_s$ becomes parallel to the one of the fundamental representation and also diquark representations $(6,10)$ get a parallel slope. Now, the string tensions of the representations which belong to the same $4$-ality class can be compared.

It was mentioned before that $Re{\cal G}_r(\alpha)$ behaves normally for the fundamental representation. It varies between $1$ and $0$ for the $z_1$ vortex and between $1$ and $-1$ for the $z_2$ one. As the same description works for each vortex type, the $z_1$ vortex is focused. Moreover, although the contribution of the $z_2$ vortex to the potential is not negligible, it is less than the contribution of the $z_1$ vortex \cite{diakonov}. Tab.~\ref{tab} shows that the string tension of representation $20_s$ at intermediate distances is larger than the fundamental one. Therefore, the center of $Re{\cal G}_r(\vec\alpha)$ passes $0$ and reaches $-0.33$ to produce an intermediate string tension larger than that of the fundamental one. Then it goes back to $0$ and the same asymptotic string tension as the fundamental one is produced.

The center of the group factor has the same value equal to $0$ for  both representations $15$ and $35_s$. However, at intermediate distances, it is equal to $0.2$ for the adjoint representation. This factor reaches $-0.25$ for representation $35_s$ so that the intermediate string tension of this representation becomes larger than the adjoint one.

For the diquark representations, $Re{\cal G}_r(\alpha)$ changes normally between $1$ and $-1$ for the $z_1$ vortex. From Tab.~\ref{tab} it is seen that the intermediate string tensions of the diquark representations are in the same range. So the group factor is expected to behave the same way for these two representations.

\begin{table}
\caption{\label{tab}Casimir scaling and $4$-ality of some SU(4) representations. the behavior of the group factor for representations with the same $4$-ality can be compared.
}
\begin{ruledtabular}
\begin{tabular}{ccccccc}
\textrm{representation}&
\textrm{4}&
\textrm{6}&
\textrm{10}&
\textrm{$15_s$(adjoint)}&
\textrm{$20_s$}&
\textrm{$35_s$}\\
\colrule
$\frac{C_r}{C_F}$ & 1 & 1.33 & 2.13 & 2.4 & 4.2 & 6.4 \\
$4$-ality & 1 & 2 & 2 & 0 & 1 & 0\\ 
\end{tabular}
\end{ruledtabular}
\end{table}

\section{\label{con}Conclusion}
The prediction of the potential between $SU(N)$ static color sources by the thick center vortices model has been very precise in various quark distances. In other words, both the intermediate and asymptotic string tensions are in agreement with the proposals given in the corresponding regions, e.g. N-ality dependence at large distances and the Sine and Casimir scaling at intermediate distances. In fact, The behavior of the string tensions depends on how the $SU(N)$ group factor interpolates between $1$ and the phase factor $\exp(\frac{2\pi i nk}{N}$) where $k$ is the N-ality of the representation and $n$ represented the vortex type. 
It is seen that the real part of the $SU(N)$ group factor changes abnormally for some representations. In the $SU(4)$ gauge group, $Re{\cal G}_r(\alpha)$ for representations $15$, $20_s$ and $35_s$, ... passes the corresponding phase factor while this factor interpolates normally between expected limits for fundamental and diquark representations. The abnormal behavior of the group factor is observed for representations $8$,$10$, $15_m$, $15_s$, ... in the $SU(3)$ gauge group. 

In this paper, I have shown that the minimum points seen in the plots of the group factor versus $x$ at large distances are located at the points where the vortex and the Wilson loop are partially linked. It means that when the position of the vortex core is placed at the minimum points, a part of the vortex flux enters the Wilson loop. The sub-Wilson loops -which make the representation $r$- affect this portion in a way that $50\%$ of the maximum flux enters the Wilson loop of the representations with N-ality=$0$. This portion turns to $60\%$ for representations which belong to N-ality=$1$ class except the fundamental representation. My calculations show that at very short distances, the center of the group factor is nearly equal to $1$ which means the vortex and the Wilson loop are slightly linked to each other. As the distance increases, this amount becomes near to $(Re{\cal G}_r(\vec\alpha))_{\mathrm{min}}$ at intermediate distances. When the quarks are separated more, the group factor center changes in a way that it gets equal to the phase factor and at this stage the asymptotic string tensions are achieved. Therefore, one might conclude that the minimum points are the positions which the group factor reaches at intermediate distances. The comparison between the asymptotic and the intermediate string tensions of the representations with the same N-ality shows that the group factor must pass the phase factor so that an intermediate string tension larger than the asymptotic one is achieved. This interpretation works for diquark representations ($6$ and $10$) which have nearly the same asymptotic and intermediate string tensions.   

\begin{acknowledgments}
I would like to thank S. Deldar for her helpful discussions. 
\end{acknowledgments}
\appendix

\section{Basis vectors of the $SU(4)$ representations}

In the following, I calculate the basis vectors for the $SU(4)$ representations $6$, $10$, $15$ $20_s$ and $35_s$.

Representation $6$:\\
This representation is made of two quarks (Eq.(\ref{diquark})). $v^i$ and $u^i$, ${i,j=1,...,4}$ are considered as the basis vectors for the quarks in fundamental representation. The young tableaux of the $6$ representation consists of two boxes in column. So the anti-symmetric basis vector might be written as:
\begin{equation}
V^{ij}=\frac{1}{2}\big(v^i u^j-v^ju^i\big).
\label{base6}
\end{equation}
It should be noticed that not all of the components of the $V^{ij}$ vectors are independent. Therefore, after omitting the additional ones, the six independent basis vectors are the following:
\begin{multline}
\begin{aligned}
X_6^1&=V^{12}&=\frac{1}{2}(v^1u^2-v^2u^1),\\
X^2_6&=V^{13}&=\frac{1}{2}(v^1u^3-v^3u^1),\\
X^3_6&=V^{14}&=\frac{1}{2}(v^1u^4-v^4u^1),\\
X^4_6&=V^{23}&=\frac{1}{2}(v^2u^3-v^3u^2),\\
X^5_6&=V^{24}&=\frac{1}{2}(v^2u^4-v^4u^2),\\
X^6_6&=V^{34}&=\frac{1}{2}(v^3u^4-v^4u^3).\\
\end{aligned}
\end{multline}

Representation $10$:\\
The same as representation $6$, two quarks make this representation but the young tableaux of representation $10$ is symmetric. Thus:
\begin{equation}
U^{ij}=\frac{1}{2}\big(v^i u^j+v^ju^i\big).
\label{base10}
\end{equation}
After removing the additional vectors, $10$ basis vectors are calculated as follows:
\begin{multline}
\begin{aligned}
X^1_{10}&=U^{11}=v^1u^1,\\
X^2_{10}&=U^{12}=\frac{1}{2}(v^1u^2+v^2u^1),\\
X^3_{10}&=U^{13}=\frac{1}{2}(v^1u^3+v^3u^1),\\
X^4_{10}&=U^{14}=\frac{1}{2}(v^1u^4+v^4u^1),\\
X^5_{10}&=U^{22}=v^2u^2,\\
X^6_{10}&=U^{23}=\frac{1}{2}(v^2u^3+v^3u^2),\\
X^7_{10}&=U^{24}=\frac{1}{2}(v^2u^4+v^4u^2),\\
X^8_{10}&=U^{33}=v^3u^3,\\
X^9_{10}&=U^{34}=\frac{1}{2}(v^3u^4+v^4u^3),\\
X^{10}_{10}&=U^{44}=v^4u^4.\\
\end{aligned}
\end{multline}

Representation $15$:\\
This representation is made of one quark and one anti-quark (Eq.(\ref{adjoint})). If $v^i$ and $u_j$, (${i,j=1,...,4}$) represent the basis vectors for quark and anti-quark respectively, the following equation shows the basis vector for the adjoint representation:
\begin{equation}
U^i_j=v^iu_j-\frac{1}{4}\delta^i_jv^ku_k.
\label{base15}
\end{equation}
After omitting the additional vectors, $15$ the basis vectors are calculated as the following:
\begin{multline}
\begin{aligned}
X^1_{15}&=U^1_1=\frac{-1}{4}(-3v^1u_1+v^2u_2+v^3u_3+v^4u_4),\\
X^2_{15}&=U^1_2=v^1u_2,\\
X^3_{15}&=U^1_3=v^1u_3,\\
X^4_{15}&=U^1_4=v^1u_4,\\
X^5_{15}&=U^2_1=v^2u_1,\\
X^6_{15}&=U^2_2=\frac{-1}{4}(v^1u_1-3v^2u_2+v^3u_3+v^4u_4),\\
X^7_{15}&=U^2_3=v^2u_3,\\
X^8_{15}&=U^2_4=v^2u_4,\\
X^9_{15}&=U^3_1=v^3u_1,\\
X^{10}_{15}&=U^3_2=v^3u_2,\\
X^{11}_{15}&=U^3_3=\frac{-1}{4}(v^1u_1+v^2u_2-3v^3u_3+v^4u_4),\\
X^{12}_{15}&=U^3_4=v^3u_4,\\
X^{13}_{15}&=U^4_1=v^4u_1,\\
X^{14}_{15}&=U^4_2=v^4u_2,\\
X^{15}_{15}&=U^4_3=v^4u_3.\\
\end{aligned}
\end{multline}

Representation $20_s$:\\
This representation consists three quarks. On the other hand, it might be produced by the tensor product of representations $4$ and $10$:
\begin{equation}
4 \otimes 10=20_s \oplus 20_m.
\label{new20}
\end{equation}
So if $V^{ij}$ and $u^i$ are the basis vectors for representations $10$ and $4$, respectively, the basis vector of representation $20_s$ might be written as follows:
\begin{equation}
W^{ijk}=\frac{1}{3}\big(V^{ij} u^k+V^{ki} u^j+V^{jk} u^i\big).
\label{base20}
\end{equation}
After omitting the additional vectors, $20$ independent basis vectors are calculated as the following:
\begin{multline}
\begin{aligned}
X^1_{20}&=X^1_{10}v^1,\\
X^2_{20}&=\frac{1}{3}(X^1_{10}v^2+2X^2_{10}v^1),\\
X^3_{20}&=\frac{1}{3}(X^1_{10}v^3+2X^3_{10}v^1),\\
X^4_{20}&=\frac{1}{3}(X^1_{10}v^4+2X^4_{10}v^1),\\
X^5_{20}&=\frac{1}{3}(X^5_{10}v^1+2X^2_{10}v^2),\\
X^6_{20}&=\frac{1}{3}(X^2_{10}v^3+X^3_{10}v^2+X^6_{10}v^1),\\
X^7_{20}&=\frac{1}{3}(X^2_{10}v^4+X^4_{10}v^2+X^7_{10}v^1),\\
X^8_{20}&=\frac{1}{3}(X^8_{10}v^1+2X^3_{10}v^3),\\
X^9_{20}&=\frac{1}{3}(X^3_{10}v^4+X^4_{10}v^3+X^9_{10}v^1),\\
X^{10}_{20}&=X^5_{10}v^2,\\
X^{11}_{20}&=\frac{1}{3}(X^5_{10}v^3+2X^6_{10}v^2),\\
X^{12}_{20}&=\frac{1}{3}(X^5_{10}v^4+2X^7_{10}v^2),\\
X^{13}_{20}&=\frac{1}{3}(X^8_{10}v^2+2X^6_{10}v^3),\\
X^{14}_{20}&=\frac{1}{3}(X^6_{10}v^4+X^7_{10}v^3+X^9_{10}v^2),\\
X^{15}_{20}&=\frac{1}{3}(X^{10}_{10}v^2+2X^7_{10}v^4),\\
X^{16}_{20}&=X^8_{10}v^3,\\
X^{17}_{20}&=\frac{1}{3}(X^{8}_{10}v^4+2X^9_{10}v^3),\\
X^{18}_{20}&=\frac{1}{3}(X^{10}_{10}v^3+2X^9_{10}v^4),\\
X^{19}_{20}&=\frac{1}{3}(X^{10}_{10}v^1+2X^4_{10}v^4),\\
X^{20}_{20}&=X^{10}_{10}v^4.\\
\end{aligned}
\end{multline}

Representation $35_s$:\\
This representation contains four quarks. On the other hand, it might be produced by the tensor product of representations $4$ and $20_s$:
\begin{equation}
4 \otimes 20_s=35_s \oplus 45.
\label{new35}
\end{equation}
So if $W^{ij}$ and $u^i$ are the basis vectors for representations $20_s$ and $4$, respectively, the basis vector of representation $35_s$ might be written as follows:
\begin{equation}
Q^{ijkl}=\frac{1}{4}\big(W^{ijk} u^l+W^{jkl} u^i+W^{kli} u^j+W^{lij} u^k \big).
\label{base35}
\end{equation}
After omitting the additional vectors, $35$ independent basis vectors are calculated as the following:

\begin{multline}
\begin{aligned}
X^1_{35}&=X^1_{20}v^1\\
X^2_{35}&=\frac{1}{4}(X^1_{20}v^2+3X^2_{20}v^1),\\
X^3_{35}&=\frac{1}{4}(X^1_{20}v^3+3X^3_{20}v^1),\\
X^4_{35}&=\frac{1}{4}(X^1_{20}v^4+3X^4_{20}v^1),\\
X^5_{35}&=\frac{1}{2}(X^2_{20}v^2+X^5_{20}v^1),\\
X^6_{35}&=\frac{1}{4}(X^2_{20}v^3+X^3_{20}v^2+2X^6_{20}v^1),\\
X^7_{35}&=\frac{1}{4}(X^2_{20}v^4+X^4_{20}v^2+2X^7_{20}v^1),\\
X^8_{35}&=\frac{1}{2}(X^3_{20}v^3+X^8_{20}v^1),\\
X^9_{35}&=\frac{1}{4}(X^3_{20}v^4+X^4_{20}v^2+2X^7_{20}v^1),\\
X^{10}_{35}&=\frac{1}{2}(X^4_{20}v^4+X^{19}_{20}v^1),\\
X^{11}_{35}&=\frac{1}{4}(X^{10}_{20}v^1+3X^5_{20}v^2),\\
X^{12}_{35}&=\frac{1}{4}(X^5_{20}v^3+2X^6_{20}v^2+X^{11}_{20}v^1),\\
X^{13}_{35}&=\frac{1}{4}(X^5_{20}v^4+2X^7_{20}v^2+X^{12}_{20}v^1),\\
X^{14}_{35}&=\frac{1}{4}(X^8_{20}v^2+2X^6_{20}v^3+X^{13}_{20}v^1),\\
X^{15}_{35}&=\frac{1}{4}(X^6_{20}v^4+X^7_{20}v^3+X^9_{20}v^2+X^{14}v^1),\\
X^{16}_{35}&=\frac{1}{4}(X^{19}_{20}v^2+2X^7_{20}v^4+X^{15}_{20}v^1),\\
X^{17}_{35}&=\frac{1}{4}(X^{16}_{20}v^1+3X^8_{20}v^2),\\
X^{18}_{35}&=\frac{1}{4}(X^{16}_{20}v^4+2X^9_{20}v^3+X^{17}_{20}v^1),\\
X^{19}_{35}&=\frac{1}{4}(X^{18}_{20}v^4+2X^9_{20}v^4+X^{19}_{20}v^3),\\
X^{20}_{35}&=\frac{1}{4}(X^{20}_{20}v^1+3X^{19}_{20}v^3),\\
X^{21}_{35}&=X^{10}_{20}v^1,\\
X^{22}_{35}&=\frac{1}{4}(X^{10}_{20}v^3+3X^{11}_{20}v^2),\\
X^{23}_{35}&=\frac{1}{4}(X^{10}_{20}v^4+3X^{12}_{20}v^2),\\
X^{24}_{35}&=\frac{1}{2}(X^{11}_{20}v^3+X^{13}_{20}v^2),\\
X^{25}_{35}&=\frac{1}{4}(X^{11}_{20}v^4+2X^{14}_{20}v^4+X^{15}_{20}v^2),\\
X^{26}_{35}&=\frac{1}{2}(X^{12}_{20}v^4+X^{15}_{20}v^2),\\
X^{27}_{35}&=\frac{1}{4}(X^{16}_{20}v^2+3X^{13}_{20}v^3),\\
X^{28}_{35}&=\frac{1}{4}(X^{13}_{20}v^4+2X^{14}_{20}v^3+X^{17}_{20}v^2),\\
X^{29}_{35}&=\frac{1}{4}(X^{15}_{20}v^3+2X^{14}_{20}v^4+X^{18}_{20}v^2),\\
X^{30}_{35}&=\frac{1}{4}(X^{20}_{20}v^2+3X^{15}_{20}v^4),\\
X^{31}_{35}&=X^{16}_{20}v^3,\\
X^{32}_{35}&=\frac{1}{4}(X^{16}_{20}v^4+3X^{17}_{20}v^3),\\
X^{33}_{35}&=\frac{1}{2}(X^{17}_{20}v^4+X^{18}_{20}v^3),\\
X^{34}_{35}&=\frac{1}{4}(X^{20}_{20}v^3+3X^{18}_{20}v^4),\\
X^{35}_{35}&=X^{20}_{20}v^4.
\end{aligned}
\end{multline}

\bibliography{final}

\providecommand{\noopsort}[1]{}\providecommand{\singleletter}[1]{#1}%
\begin{thebibliography}{10}%
\makeatletter
\providecommand \@ifxundefined [1]{%
 \ifx #1\undefined \expandafter \@firstoftwo
 \else \expandafter \@secondoftwo
\fi
}%
\providecommand \@ifnum [1]{%
 \ifnum #1\expandafter \@firstoftwo
 \else \expandafter \@secondoftwo
\fi
}%
\providecommand \enquote [1]{``#1''}%
\providecommand \bibnamefont  [1]{#1}%
\providecommand \bibfnamefont [1]{#1}%
\providecommand \citenamefont [1]{#1}%
\providecommand\href[0]{\@sanitize\@href}%
\providecommand\@href[1]{\endgroup\@@startlink{#1}\endgroup\@@href}%
\providecommand\@@href[1]{#1\@@endlink}%
\providecommand \@sanitize [0]{\begingroup\catcode`\&12\catcode`\#12\relax}%
\@ifxundefined \pdfoutput {\@firstoftwo}{%
 \@ifnum{\z@=\pdfoutput}{\@firstoftwo}{\@secondoftwo}%
}{%
 \providecommand\@@startlink[1]{\leavevmode\special{html:<a href="#1">}}%
 \providecommand\@@endlink[0]{\special{html:</a>}}%
}{%
 \providecommand\@@startlink[1]{%
  \leavevmode
  \pdfstartlink
   attr{/Border[0 0 1 ]/H/I/C[0 1 1]}%
   user{/Subtype/Link/A<</Type/Action/S/URI/URI(#1)>>}%
  \relax
 }%
 \providecommand\@@endlink[0]{\pdfendlink}%
}%
\providecommand \url  [0]{\begingroup\@sanitize \@url }%
\providecommand \@url [1]{\endgroup\@href {#1}{\urlprefix}}%
\providecommand \urlprefix [0]{URL }%
\providecommand \Eprint[0]{\href }%
\@ifxundefined \urlstyle {%
  \providecommand \doi [1]{doi:\discretionary{}{}{}#1}%
}{%
  \providecommand \doi [0]{doi:\discretionary{}{}{}\begingroup
  \urlstyle{rm}\Url }%
}%
\providecommand \doibase [0]{http://dx.doi.org/}%
\providecommand \Doi[1]{\href{\doibase#1}}%
\providecommand \bibAnnote [3]{%
  \BibitemShut{#1}%
  \begin{quotation}\noindent
    \textsc{Key:}\ #2\\\textsc{Annotation:}\ #3%
  \end{quotation}%
}%
\providecommand \bibAnnoteFile [2]{%
  \IfFileExists{#2}{\bibAnnote {#1} {#2} {\input{#2}}}{}%
}%
\providecommand \typeout [0]{\immediate \write \m@ne }%
\providecommand \selectlanguage [0]{\@gobble}%
\providecommand \bibinfo [0]{\@secondoftwo}%
\providecommand \bibfield [0]{\@secondoftwo}%
\providecommand \translation [1]{[#1]}%
\providecommand \BibitemOpen[0]{}%
\providecommand \bibitemStop [0]{}%
\providecommand \bibitemNoStop [0]{.\EOS\space}%
\providecommand \EOS [0]{\spacefactor3000\relax}%
\providecommand \BibitemShut [1]{\csname bibitem#1\endcsname}%
\bibitem{thooft}%
  \BibitemOpen
  \bibfield{author}{%
  \bibinfo {author} {\bibfnamefont{G.}~\bibnamefont{'t~Hooft}},\ }%
  \bibfield{journal}{%
  \bibinfo {journal} {Nucl.\ Phys.\ B}\ }%
  \textbf{\bibinfo {volume} {138}},\ \bibinfo {pages} {1} (\bibinfo {year}
  {1978})%
  \bibAnnoteFile{NoStop}{thooft}%
\bibitem{vin}%
  \BibitemOpen
  \bibfield{author}{%
  \bibinfo {author} {\bibfnamefont{P.}~\bibnamefont{Vinciarelli}},\ }%
  \bibfield{journal}{%
  \bibinfo {journal} {Phys.\ Lett.\ B}\ }%
  \textbf{\bibinfo {volume} {78}},\ \bibinfo {pages} {485} (\bibinfo {year}
  {1978})%
  \bibAnnoteFile{NoStop}{vin}%
\bibitem{cornwall}%
  \BibitemOpen
  \bibfield{author}{%
  \bibinfo {author} {\bibfnamefont{M.}~\bibnamefont{Cornwall}},\ }%
  \bibfield{journal}{%
  \bibinfo {journal} {Nucl.\ Phys.\ B}\ }%
  \textbf{\bibinfo {volume} {157}},\ \bibinfo {pages} {392} (\bibinfo {year}
  {1970})%
  \bibAnnoteFile{NoStop}{cornwall}%
\bibitem{feynman}%
  \BibitemOpen
  \bibfield{author}{%
  \bibinfo {author} {\bibfnamefont{R.~P.}\ \bibnamefont{Feynman}},\ }%
  \bibfield{journal}{%
  \bibinfo {journal} {Nucl.\ Phys.\ B}\ }%
  \textbf{\bibinfo {volume} {188}},\ \bibinfo {pages} {479} (\bibinfo {year}
  {1981})%
  \bibAnnoteFile{NoStop}{feynman}%
\bibitem{nielsen}%
  \BibitemOpen
  \bibfield{author}{%
  \bibinfo {author} {\bibfnamefont{H.~B.}\ \bibnamefont{Nielsen}}\ and\
  \bibinfo {author} {\bibfnamefont{P.}~\bibnamefont{Olesen}},\ }%
  \bibfield{journal}{%
  \bibinfo {journal} {Nucl.\ Phys.\ B}\ }%
  \textbf{\bibinfo {volume} {160}},\ \bibinfo {pages} {380} (\bibinfo {year}
  {1979})%
  \bibAnnoteFile{NoStop}{nielsen}%
\bibitem{ambjorn1}%
  \BibitemOpen
  \bibfield{author}{%
  \bibinfo {author} {\bibfnamefont{J.}~\bibnamefont{Ambjorn}}\ and\ \bibinfo
  {author} {\bibfnamefont{P.}~\bibnamefont{Olesen}},\ }%
  \bibfield{journal}{%
  \bibinfo {journal} {Nucl.\ Phys.\ B}\ }%
  \textbf{\bibinfo {volume} {170}},\ \bibinfo {pages} {60} (\bibinfo {year}
  {1980})%
  \bibAnnoteFile{NoStop}{ambjorn1}%
\bibitem{ambjorn2}%
  \BibitemOpen
  \bibfield{author}{%
  \bibinfo {author} {\bibfnamefont{J.}~\bibnamefont{Ambjorn}}\ and\ \bibinfo
  {author} {\bibfnamefont{P.}~\bibnamefont{Olesen}},\ }%
  \bibfield{journal}{%
  \bibinfo {journal} {Nucl.\ Phys.\ B}\ }%
  \textbf{\bibinfo {volume} {170}},\ \bibinfo {pages} {265} (\bibinfo {year}
  {1980})%
  \bibAnnoteFile{NoStop}{ambjorn2}%
\bibitem{ambjorn3}%
  \BibitemOpen
  \bibfield{author}{%
  \bibinfo {author} {\bibfnamefont{J.}~\bibnamefont{Ambjorn}}, \bibinfo
  {author} {\bibfnamefont{B.}~\bibnamefont{Felsager}},\ and\ \bibinfo {author}
  {\bibfnamefont{P.}~\bibnamefont{Olesen}},\ }%
  \bibfield{journal}{%
  \bibinfo {journal} {Nucl.\ Phys.\ B}\ }%
  \textbf{\bibinfo {volume} {175}},\ \bibinfo {pages} {349} (\bibinfo {year}
  {1980})%
  \bibAnnoteFile{NoStop}{ambjorn3}%
\bibitem{olesen}%
  \BibitemOpen
  \bibfield{author}{%
  \bibinfo {author} {\bibfnamefont{P.}~\bibnamefont{Olesen}},\ }%
  \bibfield{journal}{%
  \bibinfo {journal} {Nucl.\ Phys.\ B}\ }%
  \textbf{\bibinfo {volume} {200}},\ \bibinfo {pages} {381} (\bibinfo {year}
  {1982})%
  \bibAnnoteFile{NoStop}{olesen}%
\bibitem{deldebbio}%
  \BibitemOpen
  \bibfield{author}{%
  \bibinfo {author} {\bibfnamefont{L.}~\bibnamefont{DelDebbio}}, \bibinfo
  {author} {\bibfnamefont{M.}~\bibnamefont{Faber}}, \bibinfo {author}
  {\bibfnamefont{J.}~\bibnamefont{Greensite}},\ and\ \bibinfo {author}
  {\bibfnamefont{S.}~\bibnamefont{Olejnik}},\ }%
  \bibfield{journal}{%
  \bibinfo {journal} {Phys.\ Rev.\ D}\ }%
  \textbf{\bibinfo {volume} {55}},\ \bibinfo {pages} {2298} (\bibinfo {year}
  {1997})%
  \bibAnnoteFile{NoStop}{deldebbio}%
\bibitem{deldebbio2}%
  \BibitemOpen
  \bibfield{author}{%
  \bibinfo {author} {\bibfnamefont{L.}~\bibnamefont{DelDebbio}}, \bibinfo
  {author} {\bibfnamefont{M.}~\bibnamefont{Faber}}, \bibinfo {author}
  {\bibfnamefont{J.}~\bibnamefont{Greensite}},\ and\ \bibinfo {author}
  {\bibfnamefont{S.}~\bibnamefont{Olejnik}},\ }%
  \bibfield{journal}{%
  \bibinfo {journal} {Nucl.\ Phys.\ Proc.\ Suppl.}\ }%
  \textbf{\bibinfo {volume} {63}},\ \bibinfo {pages} {552} (\bibinfo {year}
  {1998})%
  \bibAnnoteFile{NoStop}{deldebbio2}%
\bibitem{tomboulis}%
  \BibitemOpen
  \bibfield{author}{%
  \bibinfo {author} {\bibfnamefont{T.}~\bibnamefont{Kovacs}}\ and\ \bibinfo
  {author} {\bibfnamefont{E.}~\bibnamefont{Tomboulis}},\ }%
  \bibfield{journal}{%
  \bibinfo {journal} {Nucl.\ Phys.\ Proc.\ Suppl.}\ }%
  \textbf{\bibinfo {volume} {63}},\ \bibinfo {pages} {534} (\bibinfo {year}
  {1998})%
  \bibAnnoteFile{NoStop}{tomboulis}%
\bibitem{tomboulis2}%
  \BibitemOpen
  \bibfield{author}{%
  \bibinfo {author} {\bibfnamefont{T.}~\bibnamefont{Kovacs}}\ and\ \bibinfo
  {author} {\bibfnamefont{E.}~\bibnamefont{Tomboulis}},\ }%
  \bibfield{journal}{%
  \bibinfo {journal} {Nucl.\ Phys.\ Proc.\ Suppl.}\ }%
  \textbf{\bibinfo {volume} {73}},\ \bibinfo {pages} {566} (\bibinfo {year}
  {1999})%
  \bibAnnoteFile{NoStop}{tomboulis2}%
\bibitem{reinhardt}%
  \BibitemOpen
  \bibfield{author}{%
  \bibinfo {author} {\bibfnamefont{K.}~\bibnamefont{Langfeld}}, \bibinfo
  {author} {\bibfnamefont{H.}~\bibnamefont{Reinhardt}},\ and\ \bibinfo {author}
  {\bibfnamefont{O.}~\bibnamefont{Tennert}},\ }%
  \bibfield{journal}{%
  \bibinfo {journal} {Phys.\ Lett.\ B}\ }%
  \textbf{\bibinfo {volume} {419}},\ \bibinfo {pages} {317} (\bibinfo {year}
  {1998})%
  \bibAnnoteFile{NoStop}{reinhardt}%
\bibitem{dirac}%
  \BibitemOpen
  \bibfield{author}{%
  \bibinfo {author} {\bibfnamefont{P.~A.~M.}\ \bibnamefont{Dirac}},\ }%
  \bibfield{journal}{%
  \bibinfo {journal} {Proc.\ Royal Soc.\ A}\ }%
  \textbf{\bibinfo {volume} {133}},\ \bibinfo {pages} {60} (\bibinfo {year}
  {1931})%
  \bibAnnoteFile{NoStop}{dirac}%
\bibitem{faber}%
  \BibitemOpen
  \bibfield{author}{%
  \bibinfo {author} {\bibfnamefont{M.}~\bibnamefont{Faber}}, \bibinfo {author}
  {\bibfnamefont{J.}~\bibnamefont{Greensite}},\ and\ \bibinfo {author}
  {\bibfnamefont{S.}~\bibnamefont{Olejnik}},\ }%
  \bibfield{journal}{%
  \bibinfo {journal} {Phys.\ Rev.\ D}\ }%
  \textbf{\bibinfo {volume} {57}},\ \bibinfo {pages} {2603} (\bibinfo {year}
  {1998})%
  \bibAnnoteFile{NoStop}{faber}%
\bibitem{deldar2009}%
  \BibitemOpen
  \bibfield{author}{%
  \bibinfo {author} {\bibfnamefont{S.}~\bibnamefont{Deldar}}\ and\ \bibinfo
  {author} {\bibfnamefont{S.}~\bibnamefont{Rafibakhsh}},\ }%
  \bibfield{journal}{%
  \bibinfo {journal} {Phys.\ Rev.\ D}\ }%
  \textbf{\bibinfo {volume} {80}},\ \bibinfo {pages} {054508} (\bibinfo {year}
  {2009})%
  \bibAnnoteFile{NoStop}{deldar2009}%
\bibitem{denis}%
  \BibitemOpen
  \bibfield{author}{%
  \bibinfo {author} {\bibfnamefont{D.}~\bibnamefont{Neudecker}}\ and\ \bibinfo
  {author} {\bibfnamefont{M.}~\bibnamefont{Faber}},\ }%
  \bibfield{journal}{%
  \bibinfo {journal} {PoS CONFINEMENT},\ \bibinfo {pages} {182}}%
   (\bibinfo {year} {2008})%
  \bibAnnoteFile{NoStop}{denis}%
\bibitem{deldarlat}%
  \BibitemOpen
  \bibfield{author}{%
  \bibinfo {author} {\bibfnamefont{S.}~\bibnamefont{Deldar}},\ }%
  \bibfield{journal}{%
  \bibinfo {journal} {Phys.\ Rev.\ D}\ }%
  \textbf{\bibinfo {volume} {62}},\ \bibinfo {pages} {034509} (\bibinfo {year}
  {2000})%
  \bibAnnoteFile{NoStop}{deldarlat}%
\bibitem{bali}%
  \BibitemOpen
  \bibfield{author}{%
  \bibinfo {author} {\bibfnamefont{G.~S.}\ \bibnamefont{Bali}},\ }%
  \bibfield{journal}{%
  \bibinfo {journal} {Phys.\ Rev.\ D}\ }%
  \textbf{\bibinfo {volume} {62}},\ \bibinfo {pages} {114503} (\bibinfo {year}
  {2000})%
  \bibAnnoteFile{NoStop}{bali}%
\bibitem{greensite}%
  \BibitemOpen
  \bibfield{author}{%
  \bibinfo {author} {\bibfnamefont{J.}~\bibnamefont{Greensite}}, \bibinfo
  {author} {\bibfnamefont{K.}~\bibnamefont{Langfeld}}, \bibinfo {author}
  {\bibfnamefont{S.}~\bibnamefont{Olejnik}}, \bibinfo {author}
  {\bibfnamefont{H.}~\bibnamefont{Reinhardt}},\ and\ \bibinfo {author}
  {\bibfnamefont{T.}~\bibnamefont{Tok.}},\ }%
  \bibfield{journal}{%
  \bibinfo {journal} {Phys.\ Rev.\ D}\ }%
  \textbf{\bibinfo {volume} {75}},\ \bibinfo {pages} {034501} (\bibinfo {year}
  {2007})%
  \bibAnnoteFile{NoStop}{greensite}%
\bibitem{deldar-jhep}%
  \BibitemOpen
  \bibfield{author}{%
  \bibinfo {author} {\bibfnamefont{S.}~\bibnamefont{Deldar}},\ }%
  \bibfield{journal}{%
  \bibinfo {journal} {JHEP}\ }%
  \textbf{\bibinfo {volume} {0101}},\ \bibinfo {pages} {013} (\bibinfo {year}
  {2001})%
  \bibAnnoteFile{NoStop}{deldar-jhep}%
\bibitem{deldar2005}%
  \BibitemOpen
  \bibfield{author}{%
  \bibinfo {author} {\bibfnamefont{S.}~\bibnamefont{Deldar}}\ and\ \bibinfo
  {author} {\bibfnamefont{S.}~\bibnamefont{Rafibakhsh}},\ }%
  \bibfield{journal}{%
  \bibinfo {journal} {Eur.\ Phys.\ J.\ C}\ }%
  \textbf{\bibinfo {volume} {42}},\ \bibinfo {pages} {319} (\bibinfo {year}
  {2005})%
  \bibAnnoteFile{NoStop}{deldar2005}%
\bibitem{deldar2007}%
  \BibitemOpen
  \bibfield{author}{%
  \bibinfo {author} {\bibfnamefont{S.}~\bibnamefont{Deldar}}\ and\ \bibinfo
  {author} {\bibfnamefont{S.}~\bibnamefont{Rafibakhsh}},\ }%
  \bibfield{journal}{%
  \bibinfo {journal} {Phys.\ Rev.\ D}\ }%
  \textbf{\bibinfo {volume} {76}},\ \bibinfo {pages} {094508} (\bibinfo {year}
  {2007})%
  \bibAnnoteFile{NoStop}{deldar2007}%
\bibitem{deldar2010}%
  \BibitemOpen
  \bibfield{author}{%
  \bibinfo {author} {\bibfnamefont{S.}~\bibnamefont{Deldar}}\ and\ \bibinfo
  {author} {\bibfnamefont{S.}~\bibnamefont{Rafibakhsh}},\ }%
  \bibfield{journal}{%
  \bibinfo {journal} {Phys.\ Rev.\ D}\ }%
  \textbf{\bibinfo {volume} {81}},\ \bibinfo {pages} {054501} (\bibinfo {year}
  {2010})%
  \bibAnnoteFile{NoStop}{deldar2010}%
\bibitem{deldar-g2}%
  \BibitemOpen
  \bibfield{author}{%
  \bibinfo {author} {\bibfnamefont{S.}~\bibnamefont{Deldar}}, \bibinfo {author}
  {\bibfnamefont{H.}~\bibnamefont{Lookzadeh}},\ and\ \bibinfo {author}
  {\bibfnamefont{S.~M.}\ \bibnamefont{HosseiniNejad}},\ }%
  \bibfield{journal}{%
  \bibinfo {journal} {Phys.\ Rev.\ D}\ }%
  \textbf{\bibinfo {volume} {85}},\ \bibinfo {pages} {054501} (\bibinfo {year}
  {2012})%
  \bibAnnoteFile{NoStop}{deldar-g2}%
\bibitem{georgi}%
  \BibitemOpen
  \bibfield{author}{%
  \bibinfo {author} {\bibfnamefont{H.}~\bibnamefont{Georgi}},\ }%
  \emph{\bibinfo {title} {Lie algebras in particle physics}}\ (\bibinfo
  {publisher} {Benjamin/Cummings},\ \bibinfo {address} {Massachusetts, USA},\
  \bibinfo {year} {1992})%
  \bibAnnoteFile{NoStop}{georgi}%
\bibitem{diakonov}%
  \BibitemOpen
  \bibfield{author}{%
  \bibinfo {author} {\bibfnamefont{L.}~\bibnamefont{DelDebbio}}\ and\ \bibinfo
  {author} {\bibfnamefont{D.}~\bibnamefont{Diakonov}},\ }%
  \bibfield{journal}{%
  \bibinfo {journal} {Phys.\ Lett.\ B}\ }%
  \textbf{\bibinfo {volume} {544}},\ \bibinfo {pages} {202} (\bibinfo {year}
  {2002})%
  \bibAnnoteFile{NoStop}{diakonov}%
\end{thebibliography}%

\end{document}